\documentstyle[12pt,psfig,aasms4]{article}

\received{March 26 2000}
\accepted{September 26 2000}
\begin{document}

\title{Radial density and temperature profiles of 
the intracluster gas constructed jointly from the X-ray
surface brightness measurement and the universal density profile}

\author{Xiang-Ping Wu}

\affil{Beijing Astronomical Observatory and
       National Astronomical Observatories, Chinese Academy
       of Sciences, Beijing 100012, China}

\and

\author{Tzihong Chiueh}
\affil{Department of Physics, National Taiwan University, Taipei 10746, 
    Taiwan, R.O.C.}

\begin{abstract}
In this paper we have made an attempt to derive the radial
profiles of density and temperature of intracluster gas 
based on the two well-established facts at present: 
the X-ray observed surface brightness of clusters described by 
the standard $\beta$ model and the (NFW) universal density profile 
as the underlying dark matter distribution. We have numerically solved 
the hydrostatic equation by demanding that the volumed-averaged baryon 
fraction of a cluster should asymptotically approach the universal value 
at its viral radius.  We have shown that the radial temperature 
variation derived from these constraints 
differs significantly from the conventional polytropic
equation of state: The gas temperature profile may show a dramatic increase or 
decrease with outward radius, depending sensitively on the $\beta$
parameter. A large $\beta$ value (typically $>0.8$) is required
in order to ensure that the X-ray temperature makes a drop at 
the virial radius. This indicates that 
either the NFW profile is inappropriate to apply for the overall 
dark matter distribution of clusters or other non-gravitational 
heating processes may play an important role in the dynamical evolution 
of clusters. 
\end{abstract}


\section{INTRODUCTION}

One of the major uncertainties in the cosmological applications
of dynamical properties and evolution of galaxy clusters,
e.g., the baryon fraction $f_{\rm b}$  and the X-ray luminosity-temperature
correlation, has probably  arisen from the poorly
constrained radial variations of density and especially, temperature 
of hot intracluster gas.  Indeed, the currently available X-ray spectral 
data of clusters in literature  may have suffered various instrumental
limitations (see, for example, Irwin, Bregman \& Evrard 1999), and
even conflicting  results regarding the radial temperature gradients
can be obtained by different authors based 
on different analysis techniques 
(e.g. Markevitch et al. 1998;  Irwin et al. 1999; White 2000; 
Irwin \& Bregman 2000).
This leads to an embarrassing situation that 
an oversimplified, isothermal gas distribution is often adopted 
in the estimate of the total gravitating mass $M_{\rm tot}(r)$ of a cluster
enclosed within radius $r$ 
via the hydrostatic equilibrium hypothesis.
Here, it is worthy of mentioning another possibility of
obtaining the radial distribution of intracluster gas in an indirect way, 
namely, by the inversion of the X-ray and S-Z 
observed surface brightness profiles of clusters (Silk \& White 1978; 
Yoshikawa \& Suto 1999). Yet, the actual application of 
this technique to clusters is still limited primarily 
by the current S-Z measurements.

Several efforts have thus been made towards deriving the radial profiles
of density and temperature of 
intracluster gas based on the well-motivated physical mechanisms.
Among these the shape of the gas density distribution following 
the universal density profile of the dark matter halos 
of clusters suggested by numerical simulations (Navarro, Frenk \& White 
1995; NFW) has received great interest (e.g. Makino, Sasaki \& Suto 1998; 
Ettori \& Fabian 1999; Wu \& Xue 2000a; Wu 2000; etc.). While its inner and
outer slopes are still under debate, the NFW profile may be considered to
be a good approximation of the matter distribution of dark halos. 
Alternatively, a number of high-resolution simulations have also shown
that the NFW profile is independent of mass, the initial density
fluctuation, and the cosmology (Cole \& Lacey 1996; Navarro, Frenk \& White
1997; Eke, Navarro \& Frenk 1998). Indeed, 
within the framework of isothermal and hydrostatic equilibrium
there is a striking similarity between the gas density distribution 
predicted by the NFW profile and the conventional $\beta$ model
revealed by the X-ray observations (Makino et al. 1998). 
Inclusion of the self-gravity of the gas and the polytropic equation of state 
makes it promising to directly link the theoretical predictions with
the actual X-ray observations of clusters (Suto, Sasaki \& Makino 1998).

Here, we view the problem from a different angle: Given (1)the 
radial variation of the dark matter particles characterized by the NFW profile
and (2)the X-ray observed surface brightness profiles of clusters, what can we
learn about the radial density and temperature profiles of the 
intracluster gas ?  
Namely, we begin with the two well-established `observational' facts today, and
drop the assumption about the temperature properties of the gas (isothermal 
or polytropic). Nevertheless, we include the self-gravity of the gas.
We then solve the hydrostatic equation with the following 
boundary constraint: 
The volume-averaged baryon fraction $f_{\rm b}(r)$ within the virial radius
$r_{\rm vir}$ should asymptotically approach a universal value 
$f_{\rm b,BBN}\equiv \Omega_{\rm b}/\Omega_{\rm m}$ 
defined by the Big Bang Nucleosynthesis,
where $\Omega_{\rm b}$ and $\Omega_{\rm m}$ are the average baryon 
and total mass densities of the Universe 
in units of the critical density $\rho_{\rm crit}$
for closure, respectively.
The reasons why we impose this condition on the solution
are as follows: First, it is generally believed that 
the matter composition of clusters averaged over 
a sufficiently large volume should be representative of the Universe. 
Second, the present X-ray observations have revealed an increase in
the baryon fraction of clusters with radii and no evidence for 
an asymptotic tendency towards a universal value at large radii
(White \& Fabian 1995; David 1997; White, Jones \& Forman 1997;
Ettori \& Fabian 1999; Jones \& Forman 1999; Markevitch et al. 1999).
This conflict has probably arisen from the oversimplification of
the currently adopted temperature model for the intracluster gas. 
Setting a priori
the universality of $f_{\rm b}(r_{\rm vir})$ can instead allow one to work
out the constraints on the radial variations of density and 
temperature of the gas. Third, the boundary condition will be
required mathematically in order to solve the hydrostatic equation, while 
little has been known about the gas properties within the central cores
of clusters where the local dynamical activities may play a dominant
role.

In a word, if the distribution of dark matter can be reasonably 
approximated by the NFW profile, we will be able to derive 
the radial density and temperature profiles 
of intracluster gas based on the well measured X-ray surface brightness
of clusters with the constraint that the baryon fraction within the
virial radius is universal. The resulting density and temperature
profiles can be directly compared with the X-ray spectroscopic measurements 
and the S-Z observations of clusters, which will provide
a critical test for the validity of the NFW profile and 
the hydrostatic equilibrium hypothesis. Meanwhile, such a study will be helpful
for resolving the conflict regarding the different temperature
measurements, i.e., whether or not the X-ray emitting gas can be regarded as 
isothermal in clusters. 
Eventually, we hope that this research will be of significance
for our understanding of the fundamental physical processes 
in the dynamical evolution of clusters including 
non-gravitational heating.

\section{THE MODEL}

We assume that the dark halo of a cluster is described by the NFW profile
%1
\begin{equation}
\rho_{\rm DM}(r)=\frac{\rho_{\rm s}}{(r/r_{\rm s})(1+r/r_{\rm s})^2},
\end{equation}
where $\rho_{\rm s}$ and $r_{\rm s}$ 
are the characteristic density and length, 
respectively. Another two equivalent parameters frequently 
used in literature are  $\delta_{\rm c}\equiv \rho_{\rm s}/\rho_{\rm crit}$ and
$c\equiv r_{\rm vir}/r_{\rm s}$. The virial radius $r_{\rm vir}$ is defined by
%2
\begin{equation}
M_{\rm DM}(r_{\rm vir})=\frac{4\pi}{3}r_{\rm vir}^3 
\Delta_{\rm c} \rho_{\rm crit},
\end{equation}
where $\Delta_{\rm c}$ represents 
the overdensity of dark matter with respect to
the average background value $\rho_{\rm crit}$, for which we will take 
$\Delta_{\rm c}=200$ in the following computation.  Our main conclusion is
unaffected by this choice. 
The total mass of the dark matter halo enclosed within radius $r$ is
%3
\begin{equation}
M_{\rm DM}(r)=4\pi\rho_{\rm s} r_{\rm s}^3
         \left[\ln\left(1+\frac{r}{r_{\rm s}}\right)-
          \frac{r}{r+r_{\rm s}}\right].
\end{equation}
On the other hand, the X-ray surface brightness profile $S_{\rm x}(r)$ 
can be approximated by the conventional $\beta$ model:
%4
\begin{equation}
S_{\rm x}(r)=S_0\left(1+\frac{r^2}{r_{\rm c}^2}\right)^{-3\beta+1/2},
\end{equation}
in which $r_{\rm c}$ denotes the core radius. For the thermal bremsstrahlung
emission, the above form of $S_{\rm x}(r)$ indicates 
(Cowie, Henriksen \& Mushotzky 1987)
%5
\begin{equation}
n_{\rm gas}(r)T^{1/4}(r)=n_{\rm gas,0}T_0^{1/4}
\left(1+\frac{r^2}{r_{\rm c}^2}\right)^{-3\beta/2},
\end{equation}
where $n_{\rm gas}$ and $T$ are the gas number density and temperature,
respectively. Equation (5) identifies
the $\beta$ density profile in the case of isothermality. 
As a result, the total mass in gas within $r$ is simply
%6
\begin{equation}
M_{\rm gas}(r)=4\pi\mu m_{\rm p} n_{\rm gas,0} 
\int \left(\frac{T_0}{T}\right)^{1/4}
\left(1+\frac{r^2}{r_{\rm c}^2}\right)^{-3\beta/2} r^2 dr,
\end{equation}
in which $\mu=0.585$ is the average molecular weight. 
The volume-averaged baryon fraction $f_{\rm b}(r)$ within $r$ is defined as
%7
\begin{equation}
f_{\rm b}(r)=\frac{M_{\rm gas}(r)}{M_{\rm gas}(r)+M_{\rm DM}(r)}.
\end{equation}
Here we have neglected the contribution of stellar mass to $f_{\rm b}(r)$.
We further assume that the gas is in hydrostatic equilibrium with the
underlying gravitational potential of the cluster produced by 
$M_{\rm tot}(r)\equiv M_{\rm gas}(r)+M_{\rm DM}(r)$, i.e.,
%8
\begin{equation}
\frac{G[M_{\rm gas}(r)+M_{\rm DM}(r)]}{r^2}=
-\frac{1}{\mu m_{\rm p} n_{\rm gas}}\frac{d(n_{\rm gas}kT)}{dr}.
\end{equation}
Using temperature and baryon fraction
as the two variables and re-organizing the above equations yield
%9,10
\begin{eqnarray}
\frac{d\tilde{T}}{dx}=\frac{4\beta x \tilde{T}}{x^2+a^2}-
 \frac{4\alpha_0}{3x^2}
 \left[\ln(1+x)-\frac{x}{1+x}\right]\frac{1}{1-f_{\rm b}},\\
\frac{df_{\rm b}}{dx}=\frac{(1-f_{\rm b})^2b\tilde{T}^{-\frac{1}{4}}
                     (1+\frac{x^2}{a^2})^{-\frac{3\beta}{2}}x^2
                -f_{\rm b}(1-f_{\rm b})\frac{x}{(1+x)^2}}
                {\ln(1+x)-\frac{x}{1+x}},
\end{eqnarray}
in which the scaled quantities are:
$\tilde{T}=T/T_0$, $x=r/r_{\rm s}$, $a=r_{\rm c}/r_{\rm s}$, 
$b=\mu m_{\rm p} n_{\rm gas,0}/\rho_{\rm s}$,
and $\alpha_0=4\pi G \mu m_{\rm p} \rho_{\rm s} r_{\rm s}^2/kT_0$.
Our task is thus reduced to finding the solutions of equations (9) and (10) 
with the following boundary conditions
%11
\begin{equation}
\tilde{T}(0)=1,
\end{equation}
and 
%12,13
\begin{eqnarray}
f_b(c)=f_{\rm b,BBN};\\
\frac{df_{\rm b}}{dx} \left|_{x=c}=0. \right.
\end{eqnarray}
This last constraint demands that the baryon fraction should 
asymptotically match the universal value of $f_{\rm b,BBN}$ 
at the virial radius $r_{\rm vir}$ or equivalently $c$.

\section{RESULTS}

\subsection{General properties}

There are a total of five parameters involved in equations (9) and (10): 
$\beta$, $\alpha_0$, $a$, $b$ and $\delta_{\rm c}$ or $c$. Additionally, 
the universal baryon fraction $f_{\rm b,BBN}$ should also be 
regarded as another unknown parameter although $\Omega_{\rm b}$ has been
fixed by the Big Bang Nucleosynthesis. However, these parameters are not
independent,  as we will discuss in some detail later.
Mathematically,  the number of these free parameters 
can be reduced by two with the restriction of the two first-order 
differential equations, equations (9) and (10), and the associated boundary 
conditions. Namely, given four of the six 
parameters ($\beta$, $\alpha_0$, $a$, $b$, $\delta_{\rm c}$, $f_{\rm b,BBN}$),
we will be able to find the other two from equations (9) and (10). 
Technically, we perform iteratively the numerical searches 
for the solutions of equations (9) and (10) over a space of two
parameters until the boundary conditions equations (11)-(13) are
satisfied. Using the resultant temperature $\tilde{T}$, 
we will then derive the scaled gas density 
$\tilde{n}_{\rm gas}\equiv n_{\rm gas}(r)/n_{\rm gas,0}$ 
according to equation (5):
%14
\begin{equation}
\tilde{n}_{\rm gas}=\tilde{T}^{-1/4}
 \left(1+\frac{r^2}{r_{\rm c}^2}\right)^{-3\beta/2}.
\end{equation}
We demonstrate in Figure 1 a set of typical solutions for
$a=0.3$, $\delta_{\rm c}=10^4$ and $f_{\rm b,BBN}=0.1$ but 
with different choices of $\beta$: $2/3$,  $0.75$, $0.85$ and $1$.
The other two parameters, $b$ and $\alpha_0$, 
can be uniquely determined in each case during our numerical
searches for the solutions,  which read
($b$, $\alpha_0$)=($0.34$, $5.73$), ($0.42$, $6.22$), 
($0.65$, $6.61$) and ($1.34$, $6.96$) for 
$\beta$=$2/3$,  $0.75$, $0.85$ and $1$, respectively.
The halo concentration for our example is found to be 
$c=r_{\rm vir}/r_{\rm s}=5.32$. 
We have also displayed in Figure 1 the input $S_{\rm x}(r)$ and our derived 
radial profiles of gas density, $\tilde{n}_{\rm gas}$.

\placefigure{fig1}

It is not surprising from Figure 1 that the shape of the resulting radial 
profiles of gas density, $\tilde{n}_{\rm gas}$, is very similar to 
the conventional expectation, i.e., the $\beta$ model.
Alternatively, the baryon fraction $f_{\rm b}(r)$ shows an increase with radius
roughly within the core regions, and then asymptotically approaches the
background value $f_{\rm b,BBN}$
at virial radius $c$ (for a small $\beta$) or reaches a maximum 
before it matches the background value (for a large $\beta$). 
However, the temperature profile $\tilde{T}(r)$ derived from
equations (9) and (10) along with the restrictions equations (11)--(13) 
exhibits a remarkable difference 
in the radial variation among different $\beta$ clusters, from a rapid
increase at large radii for a small $\beta$  to a sharp drop 
near the virial radius for a large $\beta$, separated roughly at 
$\beta\approx0.8$. This property is mainly due to our restriction on
the baryon fraction at $x=c$   (equations [12] and [13]): 
$f_{\rm b}=f_{\rm b,BBN}$ and $df_{\rm b}/dx=0$. Recall that a similar
temperature behavior is required in order for $f_b$
to  asymptotically approach the universal value  $f_{\rm b,BBN}$ at large radii
within the framework of the $\beta$ model and the hydrostatic 
equilibrium for intracluster gas, in which the separation of
the two different variations of the asymptotic temperature at large radii 
occurs at $\beta=2/3$ (Wu \& Xue 2000b).

Similarly, we can numerically obtain the eigenvalues ($\alpha_0$, $b$)
for different choices of ($\beta$, $a$, $\delta_{\rm c}$, $f_{\rm b,BBN}$).
In Figure 2 we display the dependence of the resultant eigenvalues 
($\alpha_0$, $b$) upon each of the four parameters.
The most interesting result is the cut-off of $\beta$ parameter set by 
$\alpha_0=0$, below which, the unphysical solutions will occur. This indicates
that the observed X-ray surface brightness profile $S_{\rm x}(r)$ described
by the $\beta$ model should not have too small values of $\beta$ if
our working scenario holds true.
In order to demonstrate how the cut-off value, $\beta_{\rm cut}$, varies
with clusters, we perform a search for the solutions with $\alpha_0=0$. 
In this case, the eigenvalue $b$ can be uniquely fixed 
by our numerical solutions. Meanwhile, our numerical search shows 
that $f_{\rm b,BBN}$ produces a minor effect
on $\beta_{\rm cut}$. Consequently, $\beta_{\rm cut}$ depends mainly on 
two parameters: $\delta_{\rm c}$ and $a$ (Figure 3). 
It turns out that for a typical cluster with  $\delta_{\rm c}=10^3$--$10^5$
and $a=0.1$--$1$, the $\beta$ parameter should be greater than 
$0.5$, and  $\beta_{\rm cut}$ may exceed $0.6$ if  $\delta_{\rm c}>10^5$.
These constraints on $\beta$ parameter seem to be in good agreement with
the X-ray observations that the majority of the X-ray clusters have 
X-ray surface brightness profiles that satisfy $\beta>0.5$. 

\placefigure{fig2}

Essentially, the parameter $\alpha_0$ depends very weakly on $\delta_{\rm c}$,
$f_{\rm b,BBN}$ and $\beta$ (for $\beta>\beta_{\rm cut}$), 
and its typical values
for $a=0.15$ and $0.3$ are around $10$ and $6.5$, respectively.
Our estimates can be compared with the previous studies that assumed 
an isothermal gas distribution tracing the gravitational potential
of the NFW profile but dropping  the restrictions of equations (12)--(13): 
For rich clusters the variation of $\alpha_0$ is limited 
to a very narrow range from $6$ to $25$ with an average value of $\sim10$ 
(Ettori \& Fabian 1999; Wu \& Xue 2000a; Wu 2000).  Actually,  $\alpha_0$ 
can be approximately regarded as a constant for a given $a$ (see Figure 2).

\subsection{Asymptotic temperature at large radii}

We learn from Figure 1 that there exists a critical value of
$\beta_{\rm crit}$ such that $\tilde{T}(c)=1$. Namely,
the radial temperature profile close to $x=c$ will show 
a tendency to increase (decrease) with radius when 
$\beta<\beta_{\rm crit}$ ($\beta>\beta_{\rm crit}$).  
The existence of such a $\beta_{\rm crit}$ can be understood qualitatively
by considering the following limiting situation where the virial radius $c$
is much greater than the gas core radius $a$.  The universal profile 
admits an $x^{-3}$ density at large $x$.  
If the distribution of $f_{\rm b,BBN}$ 
is to be asymptotically flat, the gas density 
$\tilde{n}_{\rm gas}$ must also obey $x^{-3}$ at large $x$.  Given that
the $\beta$ model requires equation (14) to hold, the existence of a
$\beta_{\rm crit}$ thus naturally follows.  For this limiting situation,
$\beta_{\rm crit}=1$ but when $c$ no longer significantly exceeds $a$, the
value of $\beta_{\rm crit}$ is found to vary somewhat as described below.

While the asymptotic high-temperature and low-temperature 
are both allowed mathematically, there should be 
one temperature profile at large cluster radii. Indeed,  
it is very unlikely that the gas temperature would 
increase monotonically with outward radius. 
At least,  there is no known mechanism   
that can produce a significant rise of gas temperature 
at outermost radii. Note that this does not exclude 
a mild increasing temperature with radius over a certain region of clusters 
(e.g. Irwin \& Bregman 2000).
As a result, the condition $\tilde{T}(c)=1$
sets another low limit on the value of $\beta$, which 
guarantees that the gas temperature would not rise at virial radius 
when $\beta>\beta_{\rm crit}$.

In order to determine $\beta_{\rm crit}$, we once again perform the 
numerical searches for the solutions of  equations (9) and (10) 
with the additional constraint $\tilde{T}(c)=1$. This last point
reduces the free parameters to three:  
$f_{\rm b,BBN}$, $\delta_{\rm c}$ and $a$. 
In Figure 3 we plot the resulting  $\beta_{\rm crit}$ against  $\delta_{\rm c}$
for two different choices of $a$ ($a=0.1$ and $a=1$) but a fixed
universal baryon fraction of $f_{\rm b,BBN}=0.1$. It appears that
although $\beta_{\rm crit}$ shows a radial variation similar to 
that of  $\beta_{\rm cut}$,  
the value of $\beta_{\rm crit}$ is apparently larger than
the corresponding result for  $\beta_{\rm cut}$, indicating that 
the observed X-ray surface brightness profiles of clusters should
possess an even larger $\beta$ parameter than $\beta_{\rm cut}$ to 
ensure the natural drop of gas temperature at the edges of clusters.
For our examples shown in Figure 3, the $\beta$ parameters are required 
to exceed $0.7$ and $0.8$ for  $\delta_{\rm c}=10^3$ and $10^4$, respectively. 
This may constitute a critical test for the working hypotheses 
established in the above section. The currently available data from
X-ray measurements seem to marginally agree with these limits
even if the effect of cooling flows (e.g. Vikhlinin et al. 1999),
and the unknown parameters 
($a$, $\delta_{\rm c}$ and $f_{\rm b,BBN}$) are taken into account.
We will explore in detail the possible reasons in the discussion section. 

\placefigure{fig3}

\subsection{Self-consistent solution}

In an optically thin plasma emission model, the central gas density
$n_{\rm gas,0}$ is related to the central surface brightness 
$S_{\rm x}(0)$ through
%15
\begin{equation}
n_{\rm gas,0}^2=4\pi^{1/2}\frac{1}{\overline{g}\;\overline{\mu}}
	 \frac{1}{\alpha(T_0)}
         \frac{\Gamma(3\beta)}{\Gamma(3\beta-1/2)}
         \frac{S_{\rm x}(0)}{r_{\rm c}},
\end{equation} 
where $\alpha(T_0)$ is the cooling function
%16
\begin{equation}
\alpha(T_0)=\left(\frac{2^4 e^6}{3m_{\rm e}\hbar c^2}\right)
         \left(\frac{2\pi kT_0}{3m_{\rm e} c^2}\right)^{1/2},
\end{equation}
$\overline{g}$ is the average Gaunt factor and 
$\overline{\mu}\equiv \mu^2 (1+X)/2$ with $X=0.768$ being the primordial
hydrogen mass fraction. The X-ray measurement of the surface brightness
profile of a cluster provides us with the following quantities, $S_{\rm x}(0)$,
$\beta$ and $r_{\rm c}$.  
On the other hand, the central electron temperature $T_0$,
can be obtained relatively easily by the X-ray spectroscopic measurements,
as compared with the outermost temperature, e.g. $T(x\approx c)$.
Given $S_{\rm x}(0)$, $\beta$, $r_{\rm c}$ and $T_0$, 
we will be able to derive the
central gas density $n_{\rm gas,0}$ in terms of equations (15) and (16). 
Moreover, since the purpose of the present paper is not to estimate
the baryon fraction of clusters, we assume that the universal value of
$f_{\rm b,BBN}$ is known. As a result, there are only two parameters,
$\rho_{\rm s}$ (or equivalently $\delta_{\rm c}$) and $r_{\rm s}$ 
in the NFW profile, 
that remain to be determined. We thus search for the solutions of
equations (9) and (10) by iterations with a set of ($\delta_{\rm c}$, $a$)
until a self-consistent solution is achieved, which uniquely determines 
the parameters $\delta_{\rm c}$ and $r_{\rm s}$ in the NFW profile. 
Recall that in the isothermal case, the X-ray surface brightness profile 
predicted by the NFW profile is directly applicable to the fitting of the
observed $S_{\rm x}(r)$,  giving rise to the two parameters,
$\delta_{\rm c}$ and $r_{\rm s}$ (Makino et al. 1998; Ettori \& Fabian 1999;
Wu \& Xue 2000a,c; Wu 2000).  
Indeed, it appears that the procedure of finding the two
parameters in the NFW profile  now becomes 
very complicated if the temperature profile of a cluster is unknown.

In the present paper we do not intend to determine the parameters 
($\rho_{\rm s}$, $r_{\rm s}$) by applying the current method to an ensemble of 
X-ray clusters. Instead, we present the result for a typical cluster
with $n_{\rm gas,0}=5\times10^{-3}$ cm$^{-3}$, 
$r_{\rm c}=0.25$ Mpc ($H_0=50$ km s$^{-1}$ Mpc$^{-1}$) 
and $T_0=7$ keV but different
$\beta$ parameters. We fix the universal baryon fraction 
to be $f_{\rm b,BBN}=0.1$.
The resulting $\rho_{\rm s}$ and $r_{\rm s}$, along with 
the concentration parameter $c$ (or equivalently, the virial radius)
and the corresponding temperature $\tilde{T}(c)$ at $x=c$,
are listed in Table 1. Apart from the significant differences in
the outermost temperatures at virial radii for different $\beta$
parameters, the values of the remaining parameters, 
$r_{\rm s}$, $\delta_{\rm c}$ and $c$, 
all fall inside the limits for typical rich clusters with an
isothermal gas distribution (Ettori \& Fabian 1999; Wu \& Xue 2000a,c; 
Wu 2000). This implies, on the other hand, that the determination of 
the two parameters in the NFW profile from the X-ray observations of
intracluster gas is probably insensitive to the radial 
temperature variations.  However, the temperature rise or drop 
near the edge of a cluster depends critically on the $\beta$ parameter.

\begin{table}
\begin{center}
\caption{$\delta_{\rm c}$ and $r_{\rm s}$ for a typical cluster$^a$}
\vskip 0.2truein
%\begin{scriptsize}
\begin{tabular}{ccccc}
\hline
$\beta$ & $r_{\rm s}$ (Mpc) & $\delta_{\rm c}$ & $c$ & $\tilde{T}(c)$ \\
\hline
0.70    & 0.51        & 7051       & 4.54& 4.66 \\
0.75    & 0.49        & 7314       & 4.63& 2.56 \\
0.80    & 0.41        & 8068       & 4.84& 2.00 \\
0.85    & 0.39        & 8600       & 4.98& 0.77 \\
1.00    & 0.29        & 11000      & 5.54& 0.06 \\
\hline
\end{tabular}
%\end{scriptsize}
\end{center}
\parbox{6.5in}{$^a$ $T_0=7$ keV, $r_{\rm c}=0.25$ Mpc and  
               $n_{\rm gas,0}=5\times10^{-3}$ cm$^{-3}$.}
 \end{table}

\section{DISCUSSION AND CONCLUSIONS}

The current X-ray measurements of the radial temperature profiles of 
intracluster gas especially at outer radii  
have some ambiguity, which is probably  responsible for the major 
uncertainties in the present determination of the gravitating masses 
of clusters. Additionally, the large dispersion of the derived
baryon fraction among different clusters may be also associated
with the poorly constrained cluster temperatures. 
Consequently, the cosmological application of the dynamical properties 
of clusters are significantly affected. In the absence of a
reliable constraint on the radial temperature variation from X-ray
spectral measurements,  we have made an attempt in the present paper 
to derive the temperature  profile $T(r)$ based on the well-established 
facts from the X-ray observations
and numerical simulations, along with some plausible boundary
conditions. Specifically, we use the NFW profile as the dark matter
distribution $n_{\rm DM}(r)$ of clusters, and assume that intracluster gas 
$n_{\rm gas}(r)$ is in hydrostatic equilibrium with the underlying 
gravitational potential dominated by  $n_{\rm DM}(r)$ and $n_{\rm gas}(r)$.
We adopt the X-ray observed surface brightness 
profile $S_{\rm x}(r)$ described by
the conventional $\beta$ model to set up a link between 
gas density  $n_{\rm gas}(r)$ and temperature  $T(r)$. 
We have then numerically 
solved the hydrostatic equation by demanding that the baryon fraction
$f_{\rm b}(r)$ approaches asymptotically the universal value 
$f_{\rm b,BBN}=\Omega_{\rm b}/\Omega_{\rm M}$ at the viral radius. As a result,
we can simultaneously determine the radial profiles of gas density, temperature
and baryon fraction in clusters. In particular, using the X-ray data of 
$\beta$, $r_{\rm c}$, $S_{\rm x}(0)$ and $T_0$, we are able to find  the
two free parameters in the NFW profile, with which the dark matter profile
can be completely fixed.

The temperature profile constructed jointly from the NFW profile 
and the X-ray surface brightness measurement appears
to be significantly different from the conventional speculation.
Apparently, our temperature profiles (see Figure 1) 
cannot be simply described by
the widely adopted polytropic equation of state, 
$T\propto n_{\rm gas}^{\gamma-1}$ 
where $\gamma$ is the polytropic index, which is reflected by 
the striking difference between the core radius of gas density and that of
temperature.  It seems that the shape of temperature profiles 
in the outer regions of clusters can be dramatically different, 
depending sensitively on the $\beta$ parameter. This point has mainly arisen  
from the restriction that the volume-averaged baryon fraction should
asymptotically approach the universal value at the virial radius.
The recent study by
Wu \& Xue (2000b) based on the $\beta$ model and the polytropic 
gas distribution  has essentially reached a  similar result,
i.e. a mild increase in temperature with radius should be required
for clusters with $1/3<\beta<2/3$ in order to ensure the universal constancy
of the cluster baryon fraction at large radii. 
Although this possibility has been reported for some clusters
by Irwin \& Bregman (2000) and shown in
Figure 1 for small $\beta$ (more clearly Figure 4 for $\beta=2/3$),  
naturally we would expect a drop in temperature at the outermost radii. 
This can be achieved if the $\beta$ parameters become larger 
(typically $\beta>0.8$). 
While numerical studies of cluster formation and evolution
often yield a similar result of  $\beta\approx0.7$--$1$,
observationally, it is unlikely that one can significantly raise 
the $\beta$ value to  $\beta>0.8$  from the fitting of the X-ray surface 
brightness profiles of clusters. Exclusion of the central cooling regions
or employment of the double $\beta$ models in the fitting of the
observed $S_{\rm x}(r)$ can only lead to a 
moderate increase of the $\beta$ parameter 
(e.g. Jones \& Forman 1984; Vikhlinin et al. 1999; Xue \& Wu 2000a).

In addition to the X-ray spectroscopic measurements, 
the SZ effect may also help to make 
a distinction between the temperature rise or fall models at large
radii close to $r_{\rm vir}$. The SZ effect depends on the gas
density and temperature via 
$\Delta T/T_{\rm CBR}\propto \int n_{\rm gas} T d\ell$,
where $T_{\rm CBR}$ represents the temperature of the cosmic background 
radiation, and the integral is performed along the line of sight $\ell$.
In Figure 4 we demonstrate the thermal SZ effect produced by two clusters
with $\beta=2/3$ and $1$, respectively, adopting the same parameters 
as in Figure 1. If the gas temperature shows a dramatic increase with radius,
the shape of the SZ profile will be strongly distorted. 
On the contrary, it turns out to be hard to 
distinguish the temperature decreasing model from the isothermal one 
simply based on the SZ observations. 

\placefigure{fig4}

Although we have fulfilled  our task that the radial profiles of 
density and temperature of intracluster gas are derived from 
the well-established facts, the existing X-ray data and 
future X-ray observations can provide a critical 
test of these predictions. Any conflict between observations and
our predictions would imply that at least one of the following working 
hypotheses should be abandoned: 
(1)the NFW profile as the dark matter distribution of clusters; 
(2)the hydrostatic equilibrium for intracluster gas; 
(3)the pure thermal bremsstrahlung as the observed X-ray emission; 
and (4)the total baryon fraction of clusters as a reliable indicator
of matter mixture of the Universe. In fact, all these hypotheses have been
challenged in recent years.

It has been noticed that the central cusp 
in the NFW profile disagrees with the soft inner mass distributions of 
galaxies and clusters,  which has led Spergel \& Steinhardt (2000) 
to propose the possibility that the cold dark matter particles 
are self-interacting. Indeed, the X-ray surface brightness of clusters
predicted from the NFW profile has a much smaller core radius than 
the one seen from X-ray observations (Makino et al. 1998).
In fact, the cusped NFW profile yields a higher central 
gas temperature, which can be clearly seen from the temperature 
profiles plotted linearly in Figure 4, while the X-ray spectroscopic
measurements often reveal a cold gas component inside the core regions.
As a result, a modification to the NFW profile
should be properly made, such as the empirical density profile with 
a definite core radius
$\rho_{\rm DM}=\rho_0(1+x)^{-1}(1+x^2)^{-1}$ (Burkert 1995), which
may both provide a good fit to the dark matter distribution 
and predict the right shape of the X-ray surface brightness
distribution of clusters (Wu \& Xue 2000c).
However, fitting the predicted X-ray surface brightness profiles
by such a revised dark halo profile to an ensemble of the X-ray observed 
surface brightness profiles of clusters, 
Wu \& Xue (2000c) have shown that the core radii of the dark matter 
halos reach the X-ray core sizes of $\sim0.2$ Mpc. The recent 
numerical simulations under some simple models for self-interacting 
dark matter particles have essentially arrived at the similar
conclusion (Yoshida et al. 2000). Such large dark matter cores apparently 
exceed the ones required by modeling giant arcs seen in the central
regions of clusters (Hammer 1991; Wu \& Hammer 1993; 
Grossman \& Saha 1994; Tyson et al. 1998).
In a sense, any attempts at significantly modifying the NFW profile may 
require novel physical mechanisms.

Alternatively, it has been argued in recent years whether the
intracluster gas is in perfect hydrostatic equilibrium with the underlying
gravitational potential,  based mainly on the presence of substructures and
complex temperature patterns in clusters revealed by optical/X-ray 
observations (e.g. Henriksen \& White 1996; Markevitch 1996).
Yet, a statistical comparison of cluster masses derived from 
optical/X-ray observations and the NFW profile, 
under the assumption of hydrostatic equilibrium,
and the strong/weak gravitational lensing measurements  
shows excellent agreement among different cluster
mass estimates on scales greater than the X-ray core radii 
(Allen 1998; Wu et al. 1998; Wu 2000; references therein). 
This suggests that the hydrostatic equilibrium
may break down only on small scales where the local dynamical activities 
make a non-negligible contribution to cluster mass estimates, and 
hydrostatic equilibrium can be reasonably applicable to 
the overall distribution of intracluster gas.

Our emphasis is thus laid on the effect of non-gravitational heating
processes in the dynamical evolution of clusters.  
It has been known in the past years that the observationally
determined X-ray luminosity ($L_{\rm x}$) and temperature relation, 
$L_{\rm x}\propto T^{3}$, deviates significantly from the prediction,
$L_{\rm x}\propto T^{2}$, under the standard scenario that assumes 
the hydrostatic equilibrium and the thermal bremsstrahlung emission for 
intracluster gas (e.g. David et al. 1993; Markevitch 1998). 
While one cannot definitely exclude the possibility 
that the baryon fraction of clusters may vary with temperature 
(David et al. 1993),  this conflict is strongly suggestive of 
the importance of the non-gravitational heating processes  
in the cosmic evolution of clusters.
In fact, supernovae in cluster galaxies would substantially inject energy into 
intracluster medium through galactic winds in the early phase of 
cluster formation, and the intracluster gas can be heated to temperatures
even exceeding that produced by gravitational heating alone
(White 1991;  David, Forman \& Jones 1991). This scenario may 
allow one to reproduce naturally the observed $L_{\rm x}$-$T$ relation 
$L_{\rm x}\propto T^{3}$ (Ponman, Cannon \& Navarro 1999;
Wu, Fabian \& Nulsen 1999; Loewenstein 2000). 
In particular,  the energy input from supernova-driven protogalactic 
winds will permit the intracluster gas to extend out to larger radii, 
giving rise to a shallower X-ray surface brightness profile
(David et al. 1990; Ponman et al. 1999; Lloyd-Davies, Ponman \& Cannon 2000). 
The discrepancy between the large $\beta$ parameters ($\beta>0.8$)
required to maintain the decreasing temperature profile near 
virial radius discussed in the present paper 
and the relatively small value of $\beta\approx0.7$ found from 
the X-ray observed surface brightness profiles of clusters 
is probably due to our negligence of the effect of other heating
processes. A reliable construction of the radial density and 
temperature profiles of intracluster gas should also allow 
the non-gravitational heating to be included.

Taking these arguments
as a whole, we feel that the exercise made in this paper, combined
with the X-ray observations of clusters, will be helpful for our
understanding of the fundamental physical process that dominates 
the dynamical evolution of clusters. Applications of the current 
method to real X-ray clusters will be presented in subsequent work
(Xue \& Wu 2000b).

\acknowledgments
We thank an anonymous referee for constructive comments that improved 
the presentation of this work.  
WXP is grateful for the hospitality of the Department of Physics
of the National Taiwan University, where part of this research was 
carried out.   This work was supported by 
the National Science Foundation of China, under Grant 1972531, and
the National Science Council of Taiwan, under Grant NSC89-2112-M008-037.

\clearpage

\begin{figure*}
\centerline{\hspace{5cm}\psfig{figure=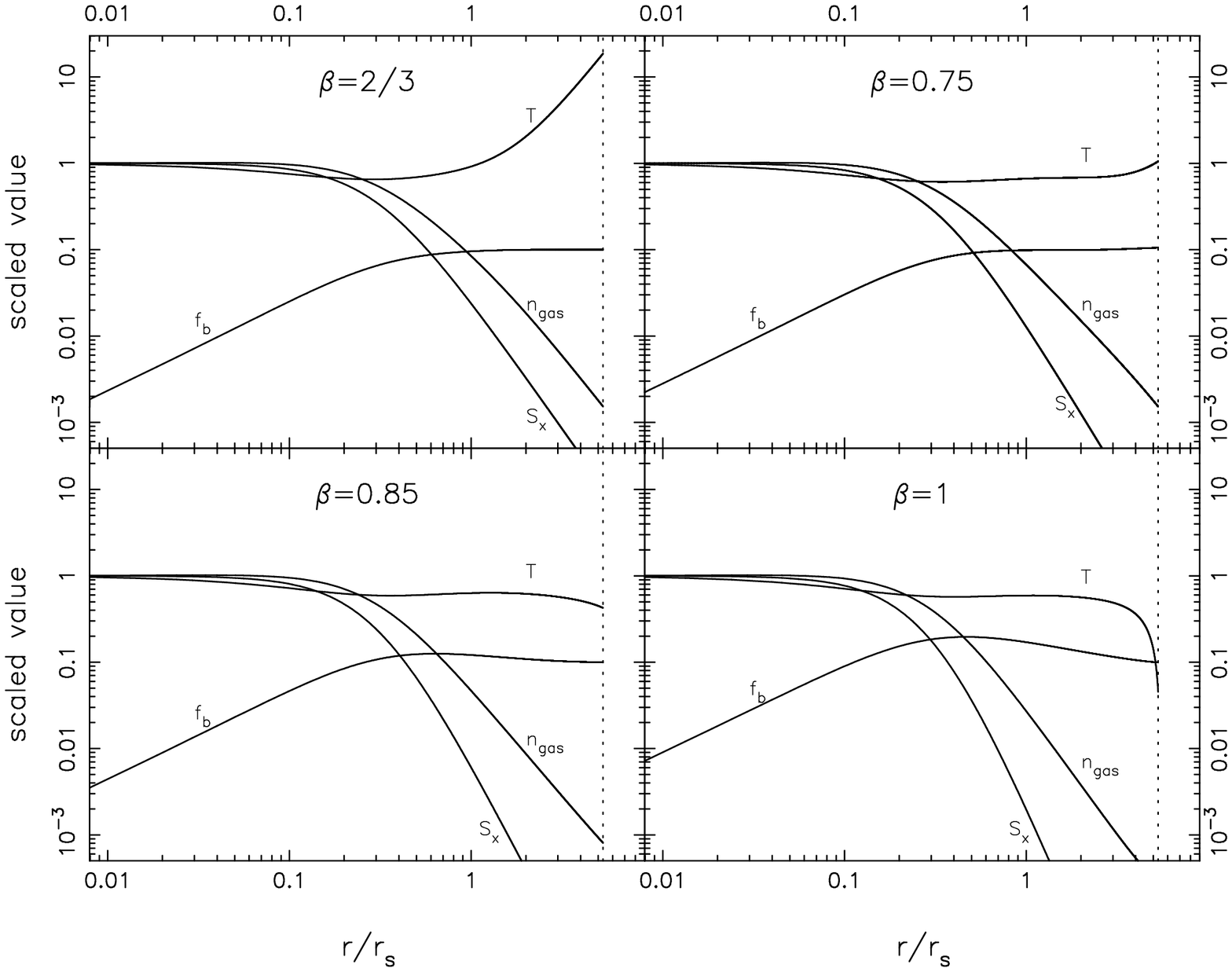,width=1.0\textwidth,angle=270}}  
\caption{The derived radial profiles of  gas density 
($n_{\rm gas}/n_{\rm gas,0}$), 
temperature ($T/T_0$) and baryon fraction ($f_{\rm b}$) for 
$f_{\rm b,BBN}=0.1$, $a=r_{\rm c}/r_{\rm s}=0.3$ 
and $\delta_{\rm c}=\rho_{\rm s}/\rho_{\rm crit}=10^4$ 
but with four different choices of $\beta$. Also shown are the input X-ray 
surface brightness profiles of clusters described by the conventional 
$\beta$ model. The dotted line denotes the virial radius. }
\end{figure*}

\begin{figure*}
\centerline{\hspace{5cm}\psfig{figure=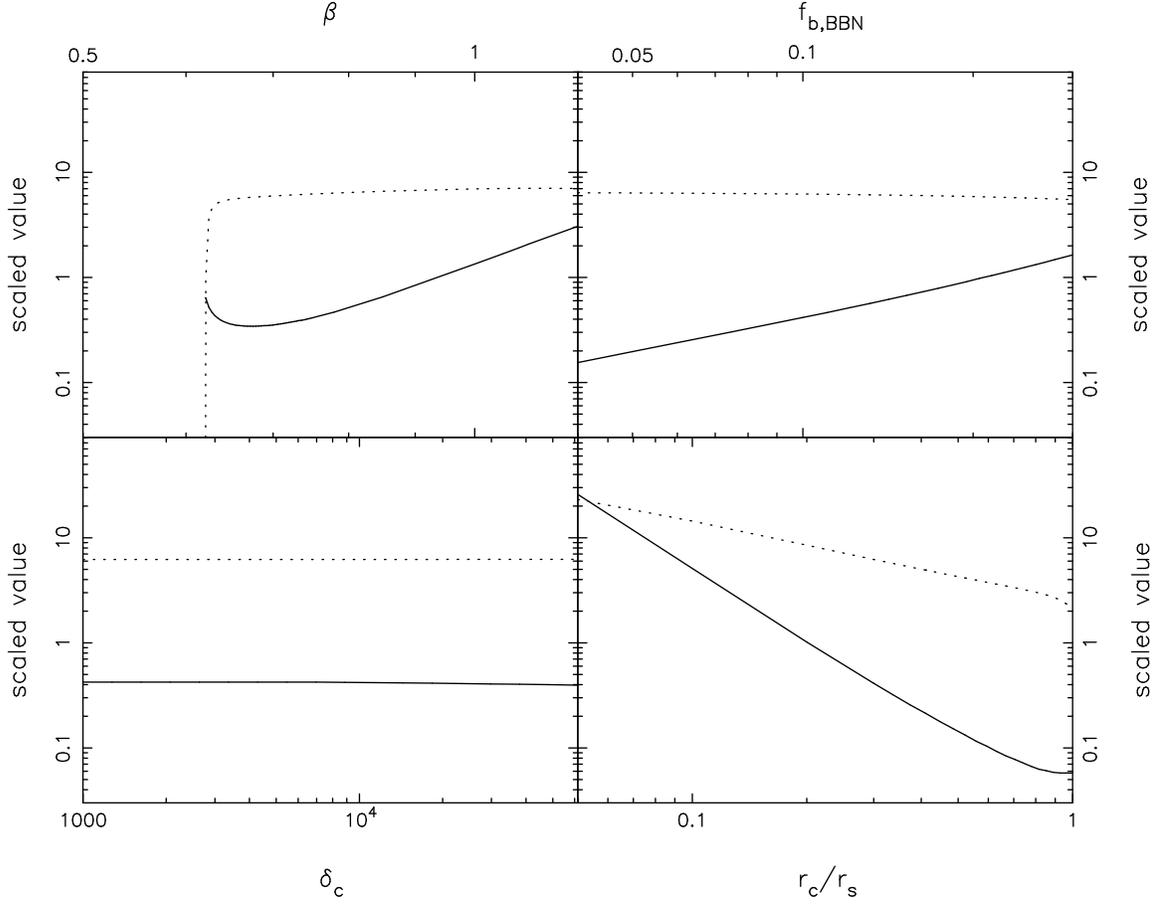,width=1.0\textwidth,angle=270}} 
\caption{Dependence of 
$\alpha_0$($=4\pi G\mu m_{\rm p} \rho_{\rm s} r_{\rm s}^2/kT_0$) 
(dotted lines) and  $b$($=\mu m_{\rm p} n_{\rm gas,0}/\rho_{\rm s}$) 
(solid lines) on 
(1)$\beta$ ($\delta_{\rm c}=10^4$, $a=0.3$ and $f_{\rm b,BBN}=0.1$), 
(2)$f_{\rm b,BBN}$ ($\beta=0.75$, $\delta_{\rm c}=10^4$ and $a=0.3$),
(3)$\delta_{\rm c}$ ($\beta=0.75$, $a=0.3$ and $f_{\rm b,BBN}=0.1$),
and (4)$a$ ($\beta=0.75$, $\delta_{\rm c}=10^4$, and $f_{\rm b,BBN}=0.1$).}
\end{figure*}

\begin{figure*}
\centerline{\hspace{5cm}\psfig{figure=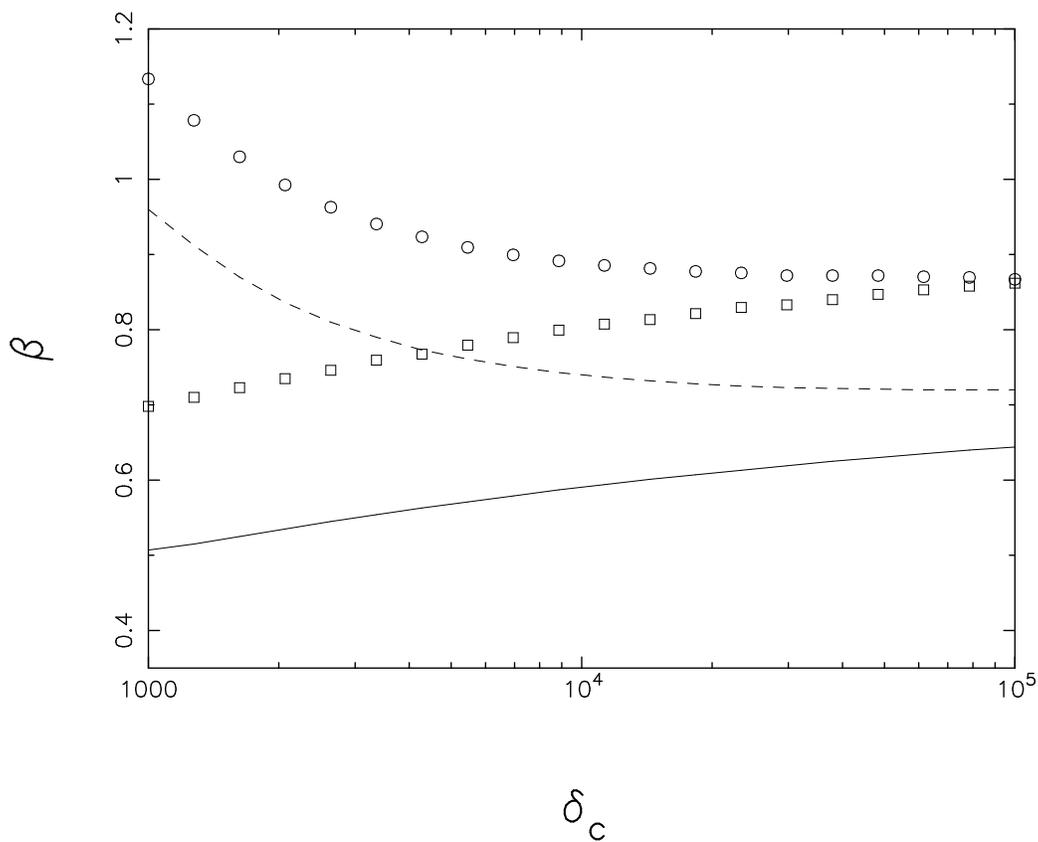,width=1.0\textwidth,angle=270}}
\caption{Lower limits on $\beta$ parameter against the characteristic 
density $\delta_{\rm c}$ for two choices of $a$: (1)$a=0.1$ (solid line and
open squares) and (2)$a=1$ (dashed line and open circles).
Solid and dashed lines: the limits $\beta_{\rm cut}$ from the unphysical 
solutions of equations (9) and (10) as a result of $\alpha_{0}=0$; 
Open squares and circles: the limits $\beta_{\rm crit}$  
set by the requirement of $T(c)/T_0=1$.}
\end{figure*}

\begin{figure*}
\centerline{\hspace{5cm}\psfig{figure=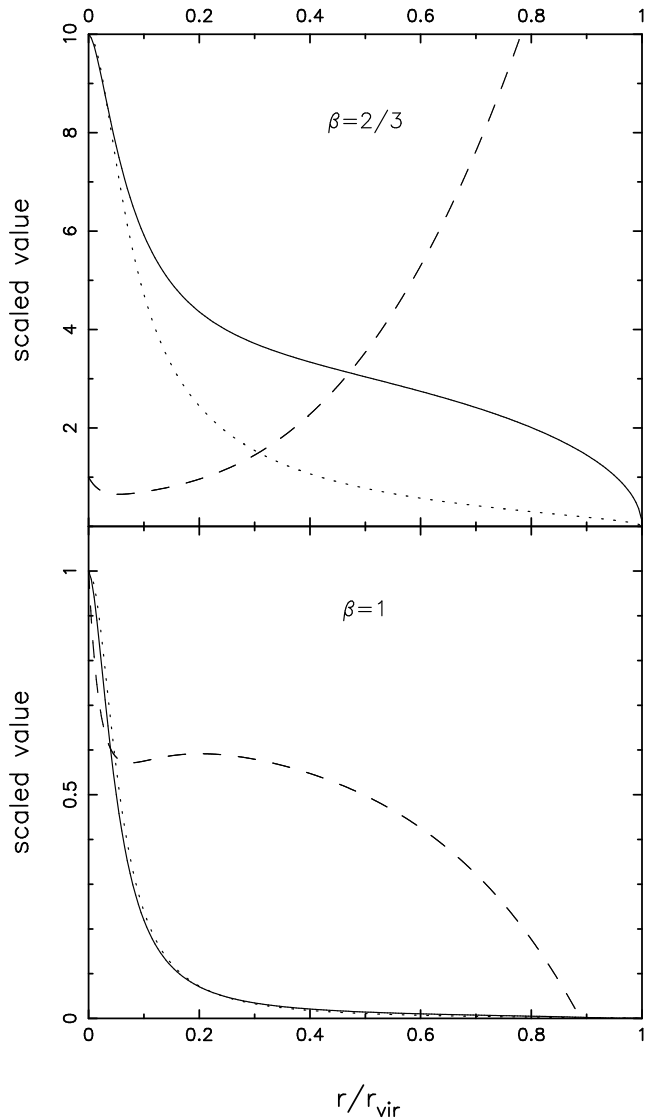,width=1.0\textwidth,angle=0}}
\caption{A comparison of the SZ profiles predicted by the isothermal 
gas distribution (dotted lines) and our derived radial distribution of 
intracluster gas from equations (9) and (10) (solid lines). Upper and 
lower panels correspond to the situations of $\beta=2/3$ and 
$\beta=1$ in Figure 1, respectively. The temperature profiles 
are shown by dashed lines. For clarity, the vertical scales for 
$\beta=2/3$ (upper panel) have been multiplied by a factor of 10.}
\end{figure*}

\end{document}